\documentclass[lettersize,journal]{IEEEtran}
\usepackage{amsmath,amsfonts}
\usepackage{algorithmic}
\usepackage{algorithm}
\usepackage{array}
\usepackage[caption=false,font=normalsize,labelfont=sf,textfont=sf]{subfig}
\usepackage{textcomp}
\usepackage{stfloats}
\usepackage{url}
\usepackage{verbatim}
\usepackage{graphicx}
\usepackage{cite}
\begin{document}

\title{Broadband tunable narrow-linewidth laser based on scattering-enhanced fiber covering E-S-C-L bands}

\author{Minzhi Xu$^1$, Zechun Geng$^1$, Da Wei, Yujia Li$^{*}$, Juntao He, Chaoze Zhang, Wei Du, Lei Gao, Leilei Shi, Ligang Huang, Jindong Wang$^{*}$, Tao Zhu$^{*}$

\thanks{Manuscript received,;revised.

This work was supported by National Key Research and Development Program of China (No. 2023YFF0715700), and the National Natural Science Foundation of China (No. 62405036; No. U23A20378; No. 625B2029), and the Exchange Project for Key Lab of Optical Fiber Sensing and Communications (Ministry of Education of China) (ZYGX2025K010). \emph{(Minzhi Xu and Zechun Geng are contributed equally to this work.) (Corresponding authors: Yujia Li; Jindong Wang; Tao Zhu.)}

Minzhi Xu, Zechun Geng, Yujia Li, Juntao He, Da Wei, Chaoze Zhang, Wei Du, Lei Gao, Leilei Shi, Ligang Huang, Jindong Wang, Tao Zhu are in Key Laboratory of Optoelectronic Technology and Systems (Education Ministry of China), Chongqing University, Chongqing 400044, China (e-mail: xu\_minzhi@163.com; gengzechun@163.com; 20220801019@stu.cqu.edu.cn; liyujia@cqu.edu.cn; 362881091@qq.com;  zhangchaoze@cqu.edu.cn; 202208131107@stu.cqu.edu.cn; Gaolei@cqu.edu.cn; shileilei@cqu.edu.cn; lghuang@cqu.edu.cn; jdwang@cqu.edu.cn; zhutao@cqu.edu.cn).
}}

\markboth{Journal of LIGHTWAVE TECHNOLOGY,~Vol.~, No.~, January~2026}%
{Shell \MakeLowercase{\textit{et al.}}: A Sample Article Using IEEEtran.cls for IEEE Journals}

\IEEEpubid{}

\maketitle

\begin{abstract}

This work demonstrates a broadband tunable narrow-linewidth laser based on scattering-enhanced fiber, covering the E-S-C-L wavelength bands from 1337.47 nm to 1631.39 nm, with a total tuning span of 293.92 nm. The laser employs two semiconductor optical amplifiers (SOAs) centered at 1420 nm and 1550 nm, which are connected into a single ring resonator via polarization multiplexing. Wavelength selection and tunability is realized using an ultra-broadband tunable filter based on a blazed grating. To suppress side longitude modes, an 18-meter-long femtosecond-laser-empowered random scattering fiber is utilized inside the cavity as a feedback medium, yielding an output linewidths between 1.54 kHz and 2.61 kHz. Benefited from the fast response of the galvanometer mirror and short relaxation time of SOAs, wavelength switching time is less than 1 ms under different tuning channels among the wavelength range of near 300 nm. The stable single-longitude-mode operation is maintained across the entire tuning range. The exceptionally broad tuning range and high spectral purity of the laser endow it with significant application potentials across a wide range of fields.

\end{abstract}


\section{Introduction}
\IEEEPARstart{H}{ing}-performance narrow-linewidth tunable lasers are widely used in applications such as optical dipole traps for atoms, high-resolution spectroscopy, and trace gas detection \cite{1,2,3,4,5,6} due to their capability of delivering single-frequency outputs at multiple wavelengths. In high-speed coherent optical communication systems, broadband tunable narrow-linewidth lasers serve as the core light source for achieving high capacity, flexibility, and reliability. Their tuning range covering the entire communication band enables flexible allocation of spectral resources in dense wavelength division multiplexing systems, facilitating wavelength-agnostic rapid service deployment and dynamic reconfiguration. Such lasers significantly simplify optical network inventory management by replacing multiple fixed-wavelength sources with a single light source.\cite{7,8,9,10}.

At present, gain media capable of providing broad spectra have been almost exhaustively explored within the field of ultrafast lasers, with their emission spectra predominantly concentrated below 1.1 $\mathrm{\mu}$m \cite{11,12,13,14}. Among these, titanium-doped sapphire is particularly renowned for its exceptionally wide gain bandwidth \cite{15,16,17}. In contrast, for the spectral regions around the optical fiber communication and eye-safe bands near 1.5 $\mathrm{\mu}$m, the available laser gain media are largely limited to erbium-doped fiber and InP-based semiconductor gain structures. Erbium-doped fiber offers a gain bandwidth of only about 60 nm, with no similar types of gain media available to seamlessly extend its spectral coverage. Semiconductor gain media, however, typically provides over 100 nm of bandwidth and can be combined with other similar gain structures, allowing for the theoretical realization of extremely broad gain spectra through seamless spectral stitching \cite{18,19,20}.

However, semiconductors exhibit a broad energy distribution during electron-hole recombination. Moreover, even minute spontaneous emission and noise within the laser cavity can induce fluctuations in carrier density, leading to rapid random variations in the resonant frequency. These factors collectively make it inherently challenging for semiconductor-based gain lasers to achieve single-longitudinal-mode operation and narrow-linewidth output. To suppress side-mode oscillation as much as possible, semiconductor lasers are often designed with ultra-short cavity lengths to maintain stable single-mode operation. This design approach, while effective for mode control, inadvertently constrains the exploitation of the intrinsic broadband gain potential of semiconductor materials \cite{21,22,23}.

To fully exploit the potential of semiconductor gain media in widely tunable lasers, researchers have continuously explored various approaches to achieve broad tuning ranges \cite{15,16,17,23,24,25,26}. Examples include the use of quantum well structures, semiconductor optical amplifier (SOA) arrays, and nonlinear frequency conversion techniques. Achieving broad wavelength tunability often involves trade-offs that can impact spectral purity. Therefore, to fulfill the dual demands of wide tuning and narrow linewidth in cutting-edge applications, the implementation of active linewidth compression mechanisms is indispensable. Common active linewidth control techniques include, but are not limited to, optical injection locking with an external cavity \cite{27,28,29}, Pound–Drever–Hall frequency stabilization \cite{30,31}, and nonlinear frequency suppression \cite{32,33}. However, these methods typically rely on fixed wavelength with high precision, making it difficultly maintain continuous wavelength tuning while simultaneously performing real-time frequency feedback control. Thus, developing a wavelength independent linewidth width compression mechanism suitable for broadband continuously tunable lasers remains one of the critical challenges to be addressed.

This work utilizes two SOAs with central wavelengths of 1420 nm and 1550 nm, which are efficiently cascaded into a single optical resonator via low loss polarization multiplexing. Combined with an ultra-broadband grating-based tunable optical filter (TOF), the laser achieves a wide tuning range from 1337.47 nm to 1631.39 nm, covering all E-S-C-L communication bands. Furthermore, an 18‑meter length of femtosecond‑laser‑empowered random scattering fiber is incorporated outside the cavity as a feedback medium to suppress frequency noise, resulting in a measured output linewidths between 1.54 kHz and 2.61 kHz.

\section{PHYSICS PRINCIPLES AND METHODS}

Fundamentally, Rayleigh scattering can occur at any physical entity, meaning that virtually any material can serve as a scatterer to participate in laser generation and noise suppression, particularly in fiber lasers and semiconductor lasers where optical fibers act as the scattering medium \cite{34,35,36}. In experiments, to enhance the collection efficiency of backscattered light and ensure compatibility with optical communication systems, the most straightforward approach is to implement this mechanism within a high-numerical-aperture optical fiber \cite{37,38}.

In a one-dimensional waveguide structure, the quasi-particles formed by a large number of randomly distributed scattering points induced by femtosecond‑laser‑empowered can be regarded as a series of random distributed-feedback planes located outside the main resonant cavity. The optical fields with random phases generated by these scattering points are injected into and coupled within the laser cavity. Therefore, based on this physical model, we start from the resonant optical field inside the main cavity. Considering a semiconductor-gain-based ring cavity, the intracavity optical field propagates unidirectionally along the Z-direction and can be expressed as:

\begin{equation}
	E_{m}(z) = E_{0} e^{ikz + \frac{1}{2}\left( g - \alpha \right) z}
\end{equation}
$g$ represents the laser gain coefficient, and $\alpha$ denotes the total loss within the semiconductor active waveguide, which includes absorption loss, scattering loss, leakage loss, and other contributions. When the output optical field from the cavity passes through the scattering-enhanced fiber, a weak scatting optical field is generated:

\begin{equation}
	E_{f}\left(z\right) = \frac{1}{\alpha_{R}} E_{0} \mathrm{e}^{ik\left(z_{i} - z\right) + \frac{1}{2}\left(g - \alpha\right)\left(z_{i} - z\right)}
\end{equation}
$\alpha_{R}$ represents the backscattering coefficient, and $z_{i}$ denotes the equivalent reflection plane position of the multi-backscattered field superposition. All single-frequency lasers should ultimately reach a steady-state operation, where the optical field mode after one round-trip in the cavity remains identical to the initial mode \cite{39}. This self-consistent condition can be expressed mathematically as:

\begin{equation}
	R\left(1 - \alpha_{\text{filter}}\right) \mathrm{e}^{2ikl + \left(g - \alpha\right) L_{c}} = 1
\end{equation}
$R$ is the coupling ratio of the ring cavity, $\alpha_{\text{filter}}$ represents the loss of the filter, and $L_{c}$ denotes the effective length of the ring cavity. The steady-state condition provides the steady‑state solution to the wave equation. Therefore, near steady state, the variation of the optical field envelope is sufficiently slow that the propagation equation can be integrated over the entire cavity length. Eventually, the rate equation for the optical field can be expressed as \cite{40,41,42}:

\begin{equation}
	\begin{aligned}
		\frac{d\widetilde{E}(t)}{dt} &= \frac{1}{2}(1 + i\alpha)G[n(t) - n_{th}] \widetilde{E}(t) \\
		&\quad + \frac{1}{\alpha_R} \sum_{i=1}^{N} \widetilde{E}_{f}(z_i) \delta(z - z_i) + \widetilde{F}_E(t)
	\end{aligned}
\end{equation}
$G$ represents the net gain of the mode, $n(t)$ denotes the carrier density, $n_{th}$ is the carrier density at threshold, $E_{f}$ is the feedback field, and the entire second term denotes the coherent sum over all feedback contributions, and $F_{E}(t)$ corresponds to the Langevin noise term.

Combining the steady-state condition of Eq. (3) and the dynamical equation of Eq. (4), we can analyze the influence of distributed backscattering feedback on the noise characteristics of the laser. Treating the feedback field $E_{f}$ as a perturbation, this perturbation introduces a phase modulation based on the main oscillating optical field in the frequency domain, equivalent to adding a distributed filtering mechanism around the laser resonance peak. Since this filtering is based on Rayleigh scattering, it exhibits no wavelength dependence. As a result, such inhomogeneity can create multiple weak Vernier effects, which adaptively assist the lasing oscillation at any central wavelength during the tuning process \cite{43,44}

\section{EXPERIMENTAL SETUP AND TUNING PERFORMANCE}

Most methods for gain interconnection employ 3 dB optical couplers (OCs) connected in parallel, which are commonly used in broadband swept-frequency lasers \cite{18,19,20}. Since OC inherently introduce 3 dB of loss, connecting two gain media with different central wavelengths in parallel would introduce 6 dB of loss, and adding more gain stages in parallel would lead to further unavoidable losses.This is disadvantageous for laser oscillation in the edge wavelength within the gain bandwidth, which limits the tuning range or flatness. In contrast, a polarization beam splitter (PBS) not only functions as a 1×2 coupler but can also theoretically reduce the loss to nearly zero by manipulating the polarization of light. This characteristic makes the PBS particularly suitable as a component for ultra-broadband tunable lasers with multiple gain media connected in parallel.

\begin{figure*}
	\centering
	\includegraphics[width=0.9 \textwidth]{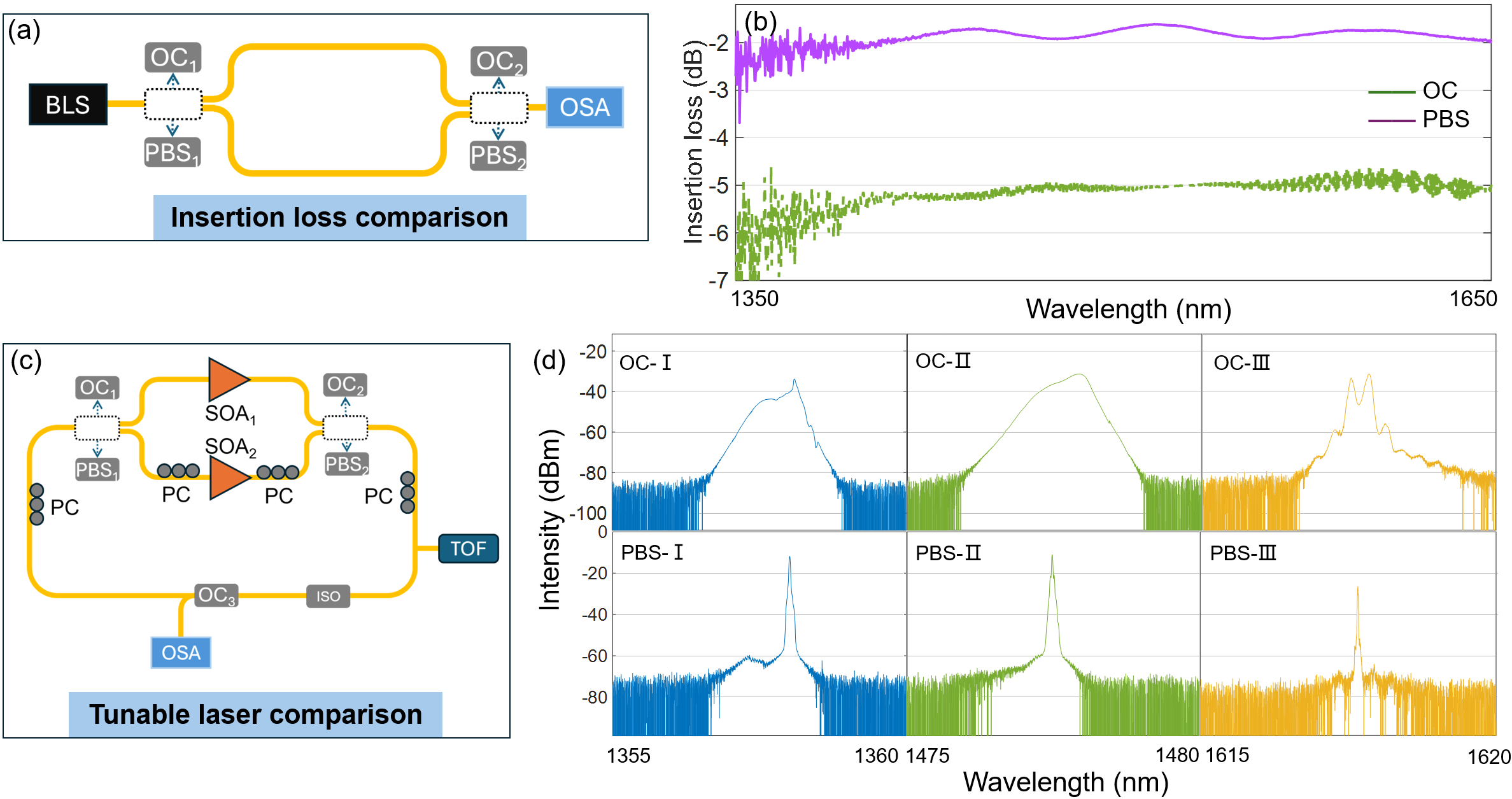}
	\caption{\label{fig:fig1} Comparative experiment between polarization beam splitting (PBS) and optical coupler (OC). (a) Experimental setup for insertion loss comparison between PBS and OC, BLS, broadband light source; OSA, optical spectrum analyzer. (b) Insertion loss of PBS versus OC. (c) Experimental setup for output laser spectrum comparison between PBS and OC. (d) Output spectra of PBS and OC.}
\end{figure*}

To better illustrate the practical insertion loss performance of the two components, a comparative experimental setup for measuring the insertion loss of PBS and OC was first constructed, as shown in Fig. 1(a). In the setup, a broadband light source (BLS) passes through port 1 of a 1×2 OC$_1$/PBS$_1$ and is split into two paths. These two paths then enter ports 2 and 3 of OC$_2$/PBS$_2$, respectively, and finally exit from port 1 of OC$_2$/PBS$_2$, which is connected to an optical spectrum analyzer (OSA). The measured insertion losses are shown in Fig. 1(b) with wavelength range of 1350-1650 nm. By adjusting the polarization at port 1, the insertion loss of the PBS can be reduced to approximately -2 dB, which is primarily attributed to the losses from the flange connections. In contrast, the OC inherently exhibits an insertion loss of -5 dB, which is twice that of the PBS.

As shown in Fig. 1(c), the PBS and OC devices are respectively connected in parallel to two SOAs operating in different wavelength bands: SOA$_1$ (BOA1410P, THORLABS) with a central wavelength of 1420 nm and SOA$_2$ (BOA1550S, THORLABS)  with a central wavelength of 1550 nm. The polarization controller (PC) handles polarization control within the ring cavity, the tunable optical filter (TOF) is responsible for tuning, and the optical isolator (ISO) ensures unidirectional laser operation. The laser output is extracted via an OC$_3$ with a splitting ratio of 2:8 and connected to an OSA. Laser spectra were measured in three wavelength regions: the 1355-1360 nm region corresponding to the gain band of SOA$_1$, the 1475-1480 nm region representing the overlap region of both SOAs, and the 1615-1620 nm region corresponding to the gain band of SOA$_2$. These are labeled as PBS/OC-\uppercase\expandafter{\romannumeral 1}, PBS/OC-\uppercase\expandafter{\romannumeral 2}, and PBS/OC-\uppercase\expandafter{\romannumeral 3}, respectively, in Fig. 1(d).

For parallel structure with OCs, laser at any wavelength generated by the two SOAs is equally split by the OC into two paths. The two parallel branches establish independent and distinct resonant conditions, leading to multi-mode oscillation. Moreover, cross absorption between SOAs intensifies mode competition and instability, resulting in severely broadened spectra as shown for OC-\uppercase\expandafter{\romannumeral 1}, OC-\uppercase\expandafter{\romannumeral 2}, and OC-\uppercase\expandafter{\romannumeral 3} in Fig. 1(d). By adjusting the polarization state, the PBS forces the resonant laser fields from the two different SOAs to become orthogonally polarized. This prevents mutual interference between the modes excited by the two SOA gains, allowing each to propagate independently. As a result, single-longitudinal-mode lasing is more readily achieved. The corresponding output spectra are presented as PBS-\uppercase\expandafter{\romannumeral 1}, PBS-\uppercase\expandafter{\romannumeral 2}, and PBS-\uppercase\expandafter{\romannumeral 3} in Fig. 1(d).

Based on the above discussion, the designed experimental setup of the broadband tunable narrow-linewidth laser is illustrated in Fig. 2(a). Two polarization beam splitters, PBS$_{1}$ and PBS$_{2}$, are used to combine the two SOAs with different central wavelengths in parallel. Since both the spontaneous emission and the laser gain of the SOAs are predominantly TE-polarized, in order to orthogonally polarize the outputs from the two SOAs, the output arm of SOA$_{1}$ is aligned with the TE-polarized arm of PBS$_{1}$ and PBS$_{2}$ via polarization-maintaining fiber. In contrast, the output arm of SOA$_{2}$, constructed using single-mode fiber, is aligned with the TM-polarized arm of PBS$_{1}$ and PBS$_{2}$ by a compact polarization controller (CPC). Furthermore, polarization controllers PC$_{1}$ and PC$_{2}$ are inserted into the cavity to independently adjust the polarization state entering the TOF and the polarization injected into PBS$_{1}$, respectively. The TOF is composed of a fiber collimator (FC), a galvanometer mirror (GM), and a blazed grating (BG) with a line density of 600 lines/mm. The GM employs a piezoelectric transducer-driven off-axis design. The filter is configured as a reflective type via an optical circulator-1 (CIR$_{1}$). An optical ISO ensures unidirectional operation within the ring cavity. The output OC has a splitting ratio of 2:8, with the 80\% output port connected to port 1 of CIR$_{2}$. A scattering-enhanced fiber (SEF) is connected to Port 2 of CIR$_{2}$ along with a fiber optic reflector (FOR). Within the SEF, the backscattered laser generated by the two counter-propagating optical fields exhibits a relative phase difference that fully covers the entire range from 0 to 2$\pi$, which is then fed back into the ring cavity.

\begin{figure*}
	\centering
	\includegraphics[width=0.9 \textwidth]{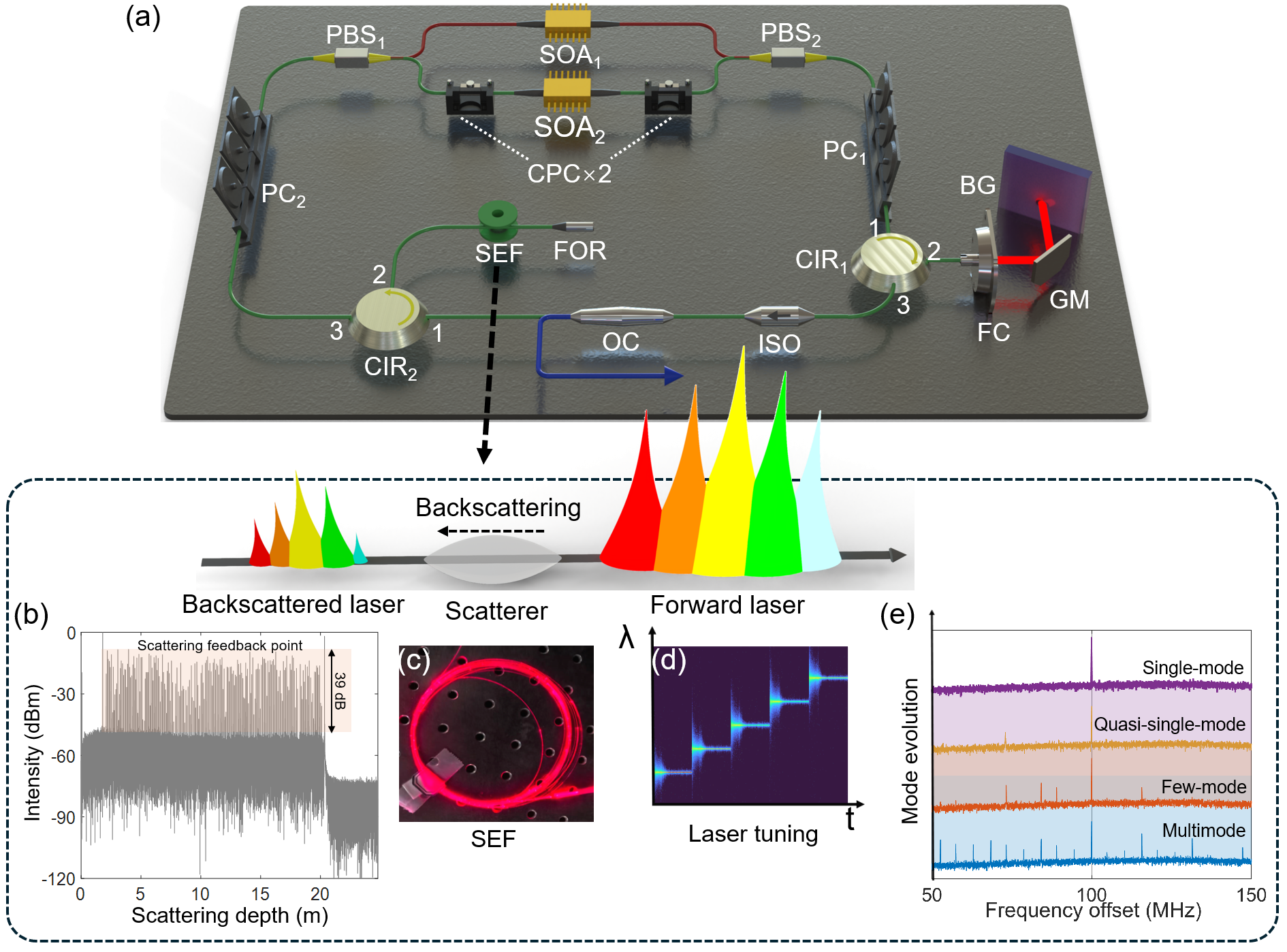}
	\caption{\label{fig:fig2} Experimental setup and schematic diagram. (a)  Experimental setup: SOA$_{1}$ and SOA$_{2}$, semiconductor optical amplifier; PBS$_{1}$ and PBS$_{2}$, polarization beam splitter; CPC, compact polarization controller; PC$_{1}$ and PC$_{2}$, polarization controller; CIR$_{1}$ and CIR$_{2}$, circulator; FC, fiber collimator; GM, galvanometer mirror; BG, blazed grating; ISO, isolator; OC, optical coupler; SEF, scattering-enhanced fiber; FOR, fiber optic reflector. (b) Backward optical signal measurement via scattering-enhanced fiber. (c) scattering-enhanced fiber image. (d) Laser wavelength tuning schematic. (e) Characterization of mode evolution via SOA current tuning.}
\end{figure*}

As shown in the inset of Figs. 2(b) and (c), the SEF is fabricated by femtosecond-laser-induced random modification of an 18 m high-numerical-aperture fiber and measurement of reflected signals. This scattering fiber acts as a random scattering feedback element, which assists the laser in suppressing side-mode oscillation and achieving narrow-linewidth output. Figure 2(b) displays the optical frequency-domain reflectometry trace measured from the SEF, where the irregular peaks correspond to randomly distributed scattering points. After the femtosecond‑laser‑empowered scattering treatment, the backscattering intensity is approximately 39 dB higher than that of a conventional high-numerical-aperture fiber. Figure 2(d) schematically provides the  process of the linewidth narrowing in different tuning channels. The proposed structure can be employed at any wavelength within the tuning range, provided that sufficient scattering is induced in the waveguide. By gradually adjusting the output power of the SOA, the feedback strength from the SEF is varied. As can be seen from Fig. 2(e), the laser output evolves progressively from multi-longitudinal mode operation to a few-mode operation, then to a quasi-single-mode operation, and finally to a stable single-mode output.

Here, the reflective filter is a home-made Littrow-configuration TOF with a wide tuning range, whose detailed structure is shown in Fig. 3(a). To ensure uniform imaging quality across the entire tuning range of nearly 300 nm, an achromatic lens is used to collimate the output beam from the fiber. The lens has a focal length of f = 11.2 mm. Wavelength tuning is achieved by electrically rotating GM to change the reflection angle, thereby selecting different wavelengths for resonance. The measured data in Fig. 3(b) show the normalized intensity reflection spectrum over the 1260-1700 nm band, indicating the wavelength independence of the filter reflectivity and demonstrating that residual lens chromatic aberration is largely eliminated. The figure indicates that the increase in insertion loss around 1200 nm is due to the operational bandwidth limitation of the circulator, while beyond 1700 nm, the behavior is primarily constrained by the measurement range limit of the optical spectrum analyzer. Based on measurements of 88 tunable reflection spectra, the measured insertion loss lies between -5.49 $\pm$ 0.41 dB, with a 3 dB bandwidth of 2.01 $\pm$ 0.06 nm as presented in Fig. 3(c). Losses greater than -7 dB persist at the edge wavelengths, which is attributed to alignment inaccuracies in the manually calibrated geometric optical path.

Prior to parallelizing the SOAs, the spontaneous emission spectra of the two SOAs were measured individually, as shown in Fig. 3(d). Their 10 dB bandwidths are 1346.2-1479.6 nm and 1468.8-1624.6 nm, respectively, with an overlapping spectral region of 1468.8-1479.6 nm. By polarization-multiplexing and cascading the two SOAs, an ultra-wide tuning range of 1337.5-1631.4 nm is achieved, corresponding to a total tuning span of 293.9 nm and a tuning step of 12 pm through precise drive voltage adjustment, as illustrated in Figs. 3(e) and (f).

\begin{figure*}
	\centering
	\includegraphics[width=0.9 \textwidth]{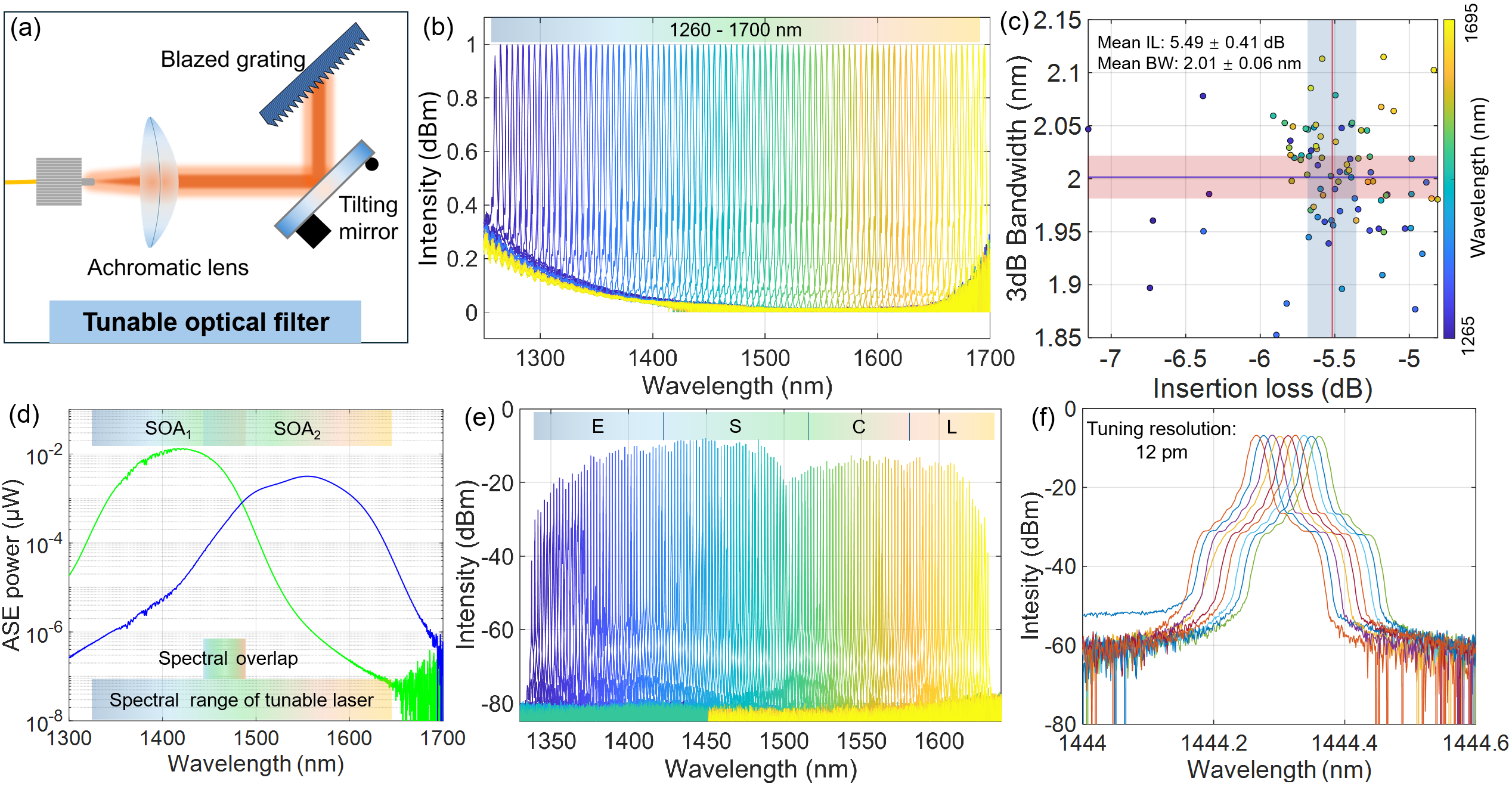}
	\caption{\label{fig:fig3}  Characterization of the tunable filter and laser.  (a) Designed structure of the tunable optical filter. (b) Tunable reflection spectra of the optical filter. (c) Statistical analysis of optical filter insertion loss and 3 dB bandwidth at different wavelengths tuning. (d) Spontaneous emission spectra of SOA$_{1}$ and SOA$_{2}$. (e) Optical spectra of the tunable laser. (f) Precisely tuned laser spectra.}
\end{figure*}

\section{TIME‑FREQUENCY CHARACTERIZATION AND DISCUSSION}

\begin{figure*}
	\centering
	\includegraphics[width=0.6 \textwidth]{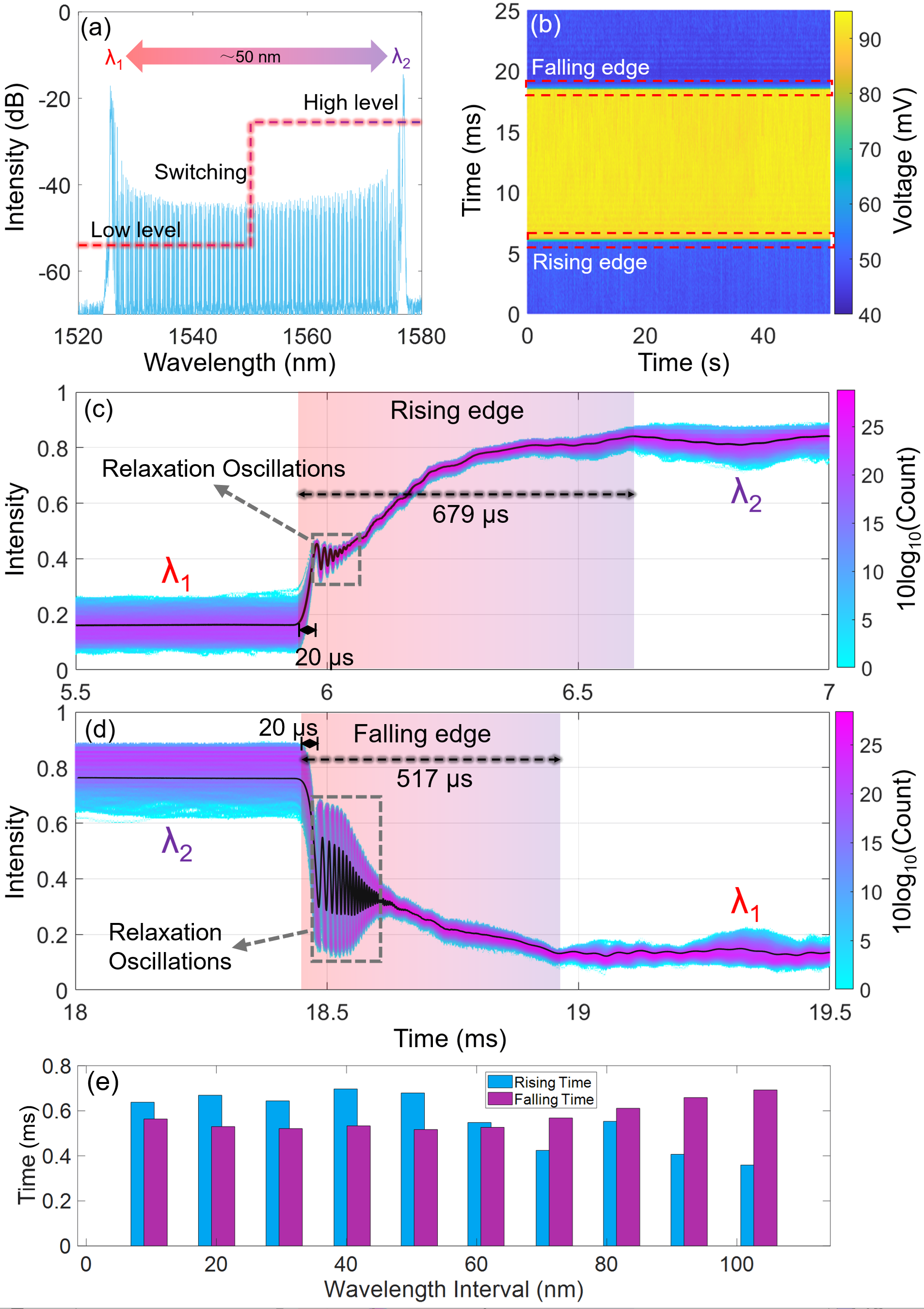}
	\caption{\label{fig:fig4} Analyzing and evaluating the dynamic behavior and process of wavelength switching. (a) Integrated spectral profiles during wavelength switching. (b) Concatenated laser intensity signals representing wavelength switching. (c) Eye diagram of temporal evolution for switching from $\lambda_{1}$ to $\lambda_{2}$. (d) Eye diagram of temporal evolution for switching from $\lambda_{2}$ to $\lambda_{1}$. (e) Statistics of switching interval durations.}
\end{figure*}

The tuning speed, and linewidth of laser represent higher-order performance dimensions and are among the most critical metrics in practical applications. Therefore, characterizing these two parameters is essential. We first examine the dynamic switching behavior between 1525 nm and 1575 nm, i.e., from $\lambda_{1}$ to $\lambda_{2}$. Fig. 4(a) shows the spectral evolution recorded by an optical spectrum analyzer during wavelength switching. Fig. 4(b) presents a time-domain intensity map of the switching process captured by an oscilloscope. The GM of the filter is driven by a 40 Hz square-wave signal. The accumulated intensity signals over a single period of 50 s are displayed in Fig. 4(b), where the blue regions correspond to the low signal level ($\lambda_{1}$) and the yellow regions to the high signal level ($\lambda_{2}$). The transitions between blue and yellow represent the rising and falling edges, corresponding to the switching from $\lambda_{1}$ to $\lambda_{2}$ and from $\lambda_{2}$ to $\lambda_{1}$, respectively. The Y-axis side-view eye diagrams of these transitions are shown in Figs. 4(c) and (d).

Fig. 4(c) illustrates the wavelength switching process from $\lambda_{1}$ to $\lambda_{2}$, with a measured switching time of 679 $\mathrm{\mu}$s. The plot clearly reveals that the re-establishment of the laser wavelength undergoes a rapid and large intensity change lasting about 20 $\mathrm{\mu}$s, which corresponds to the time required for the GM to complete the wavelength switching. The process of establishing a new wavelength commences with relaxation oscillations that persist for several hundred microseconds, with an oscillation frequency on the order of tens of kHz. As shown in the figure, these oscillations are primarily ascribed to the superposition of residual GM vibrations and the laser's intrinsic relaxation dynamics. Subsequently, the laser oscillation gradually stabilizes. The switching process from $\lambda_{2}$ to $\lambda_{1}$ is similar, as shown in Fig. 4(d). The total switching times for the two directions are 679 $\mathrm{\mu}$s and 517 $\mathrm{\mu}$s, respectively, demonstrating the fast tuning capability of the laser.

The wavelength switching processes for different step sizes centered at 1550 nm were also statistically analyzed. Switching times for step sizes ranging from 10 nm to 100 nm remain around 0.6 ms. As observed in Fig. 4(e), for step sizes between 10 nm and 60 nm, the switching time from $\lambda_{1}$ to $\lambda_{2}$ is longer than that from $\lambda_{1}$ to $\lambda_{2}$; for step sizes between 60 nm and 100 nm, the opposite trend is observed. The tunable laser comprises an SOA gain section and a fiber cavity, establishing a new lasing mode requires a certain number of cavity round‑trips, which explains why the wavelength switching time is significantly longer than that of a monolithic semiconductor laser. It is anticipated that further shortening the cavity length would correspondingly reduce the switching time.

\begin{figure*}
	\centering
	\includegraphics[width=0.9 \textwidth]{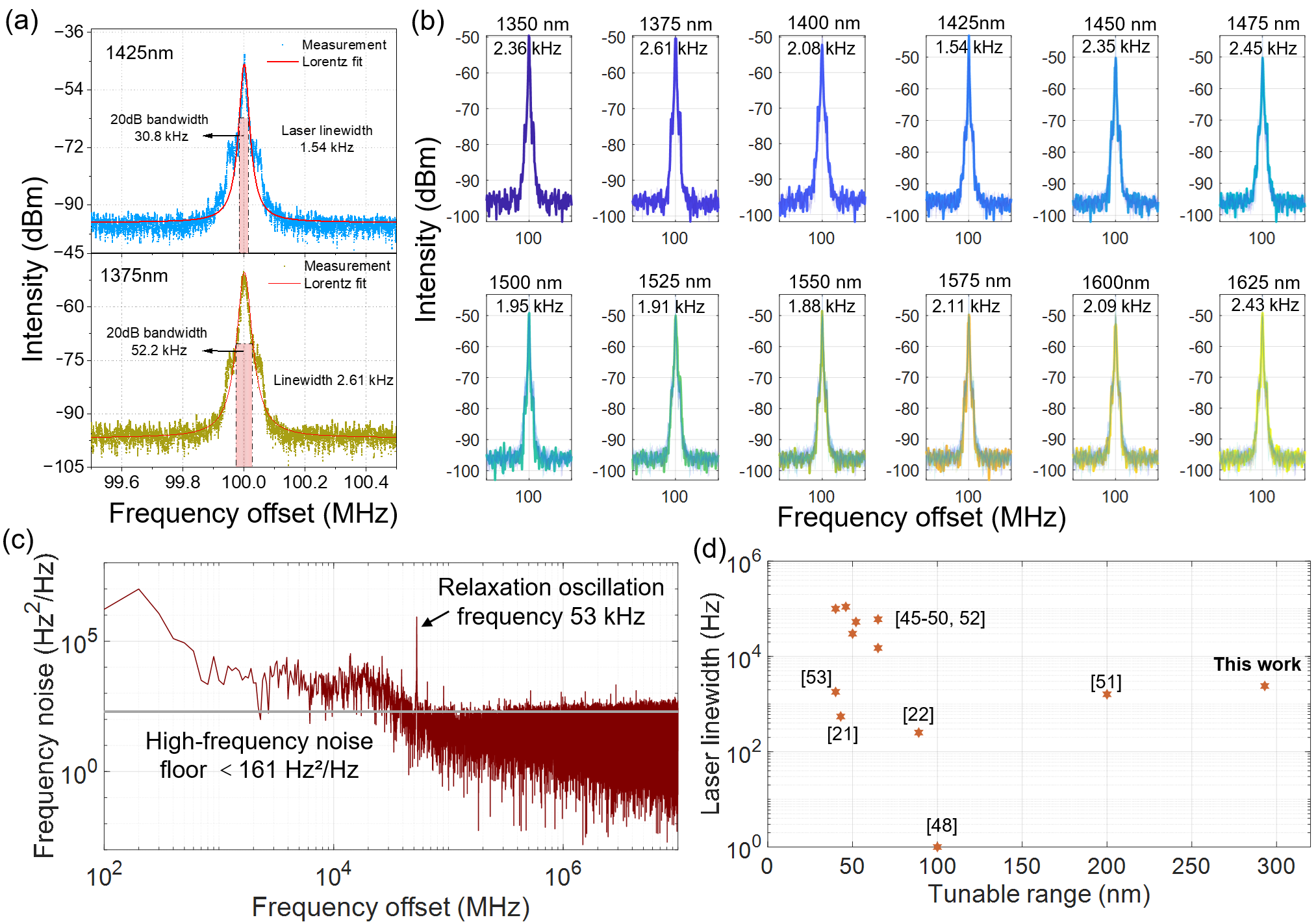}
	\caption{\label{fig:fig5} Characterization of tunable laser frequency characteristics. (a) Laser linewidths at different wavelengths tuning. (b) Concatenated laser linewidths across wavelengths. (c) Frequency noise at different wavelengths tuning. (d) Performance comparison with the state of the art and previous research.}
\end{figure*}

The laser linewidth was measured at different wavelengths tuning, as shown in Figs. 5(a) and (b). The laser linewidth across the 1350-1625 nm range was characterized using the delayed self-heterodyne interferometry method with a 50-km delay line. The measured 20 dB bandwidth extremes were 30.8 kHz and 52.2 kHz, corresponding to laser linewidths of 1.54 kHz and 2.61 kHz, respectively, as shown in Fig. 5(a). A pronounced relaxation-oscillation peak near 50 kHz is clearly visible in the beat-note envelope, which aligns well with the measurements presented in Figs. 4(c) and (d). Figure (b) presents the measured linewidth values across the wavelength range of 1350-1625 nm. The data generally exhibits the characteristic trend where the linewidth reaches its minimum at the central wavelengths of the two SOAs and increases towards the spectral edges. The frequency noise at different wavelengths tuning was characterized by coherent delayed self-heterodyne interferometry, shown in Fig. 5(c). The high‑frequency noise floor generally is less than 161 Hz$\mathrm{^{2}}$/Hz, corresponding to an intrinsic linewidth of less than 506 Hz. This increase is partly attributed to the lower laser power at the tuning extremes, which reduces the signal-to-noise ratio of the acquired signal and consequently enlarges the measurement uncertainty of the high-frequency noise floor. A pronounced relaxation oscillation peak is observed at approximately 53 kHz, with a peak level of about 916k Hz$^2$/Hz. This peak corresponds to the characteristic frequency of carrier-photon interaction within the laser cavity and is closely related to the injection current of the SOA and the intracavity loss. In the low-frequency region, the frequency noise spectrum exhibits typical 1/f behavior. Due to the absence of active frequency stabilization, the low-frequency noise is significantly influenced by factors such as SOA current noise, thermal fluctuations, and environmental acoustic and vibrational disturbances.

Figure 5(d) presents a comprehensive performance comparison between the laser developed in this work and recently reported broadband tunable narrow-linewidth lasers, focusing on two key metrics, tuning range and spectral linewidth. Each scatter point in the plot corresponds to a result from the literature. As can be seen, most previously reported lasers are clustered in the region with a tuning range below 100 nm, reflecting the long-standing trade‑off between wide tunability and narrow linewidth. In contrast, this work achieves a wavelength tuning span of 293.9 nm while maintaining a linewidth on the order of 2.6 kHz, demonstrating overall performance that surpasses existing studies.

\section{CONCLUSION}

This study effectively addresses the fundamental trade-off between spectral coverage and laser linewidth in semiconductor-based tunable sources through a laser architecture that combines polarization-multiplexed parallel gain with scattering-enhanced distributed feedback. The implementation of a femtosecond‑laser‑empowered randomly scattering fiber enables a wavelength-independent adaptive feedback mechanism, achieving universal linewidth compression across an ultra-wide tuning range. Experimental characterization confirms stable single-frequency operation over a 293.9 nm bandwidth, while maintaining sub-millisecond switching dynamics and a laser linewidth below 2.61 kHz. This design provides a valuable reference for the development of broadband tunable lasers.

\vfill


\begin{thebibliography}{1}
\bibliographystyle{IEEEtran}

\bibitem{1}Greiner, M., Mandel, O., Esslinger, T. et al. "Quantum phase transition from a superfluid to a Mott insulator in a gas of ultracold atoms", \textit{Nature}, vol. 415, pp. 39--44, 2002.

\bibitem{2}M. Aidelsburger, M. Atala, M. Lohse, et al., "Realization of the Hofstadter Hamiltonian with Ultracold Atoms in Optical Lattices",  \textit{Phys. Rev. Lett.}, vol. 111, no. 185301, 2013.

\bibitem{3}Y. Dong, W. Luo, W. Li, et al., "Focus on sub-nanometer measurement accuracy: distortion and reconstruction of dynamic displacement in a fiber-optic microprobe sensor", \textit{Light Adv. Manuf.}, vol, 5, no. 51, 2024.

\bibitem{4}Y. Chen; J. Xia; R. Zhang, et al., "Machine Learning-Enhanced Laser Absorption Spectroscopy for Harsh-Environment Combustion Diagnosis", \textit{IEEE Trans. Instrum. Meas.}, vol. 74, pp. 1--10, 2025.

\bibitem{5}Y. Cheng, Y. Xu, T. Chen, et al., "Differential laser-induced thermoelastic spectroscopy for dual-gas CO2/CH4 detection", \textit{Measurements}, vol. 240, no. 1, no. 115594,, 2025.

\bibitem{6}L. Dong, Y. Yu, C. Li, et al., "Ppb-level formaldehyde detection using a CW room-temperature interband cascade laser and a miniature dense pattern multipass gas cell", \textit{Opt. Express },
vol. 23, no. 2, pp. 19821--19830, 2015.

\bibitem{7}M. Earnshaw, C. Bolle, R. Kopf, et al., "Compact, efficient ultra-wide tunable laser with reduced thermal crosstalk", \textit{Opt. Express}, vol. 32, pp. 33283--3329, 2024.

\bibitem{8}H. Al-Taiy, N. Wenzel, S. Preußler, et al., "Ultra-narrow linewidth, stable and tunable laser source for optical communication systems and spectroscopy", \textit{Opt. Lett}, vol. 39, pp. 5826--58291, 2014.

\bibitem{9}Z. Ruan, Y. Wan, L. Wang, et al., "Flexible orbital angular momentum mode switching in multimode fibre using an optical neural network chip", \textit{Light Adv. Manuf.} vol. 5, no. 23, 2024.

\bibitem{10}Q. Sun, L. Wang, M. Zheng, et al., "Demonstration of all-optical switching system based on ultrafast tunable lasers based on DFB laser arrays", \textit{Proc. SPIE, Semiconductor Lasers and Applications XIII}, no. 1276115, 2023. 

\bibitem{11}X. Li, M. Reber, C. Corder, et al., "High-power ultrafast Yb:fiber laser frequency combs using commercially available components and basic fiber tools", \textit{Rev. Sci. Instrum.}, vol. 87, no. 093114, 2016.

\bibitem{12}U. Demirbas, "Cr:Colquiriite Lasers: Current status and challenges for further progress,
Progress in Quantum Electronics", \textit{Prog. Quantum Electron.}, vol. 68, no. 100227, 2019.

\bibitem{13}T. Ma, Y. Yao, J. Zhang, et al., "High-power high-beam-quality picosecond laser", \textit{Opt. Laser Technol.}, vol. 184, no. 112567, 2025.

\bibitem{14}Coherent. (2025, December 15). Ultrashort Pulse (USP) Lasers. Retrieved December 26, 2025, from https://www.coherent.com/lasers/ultrashort-pulse.

\bibitem{15}J. Wei, X. Cao, P. Jin, et al., "Realization of compact Watt-level single-frequency continuous-wave self-tuning titanium: sapphire laser", \textit{Opt. Express}, vol. 29, pp.  2679--2689, 2021.

\bibitem{16}C. Peng, X. Liang, R. Liu, et al., "Two-beam coherent combining based on Ti:sapphire chirped-pulse amplification at the repetition of 1 Hz", \textit{Opt. Lett.}, vol. 44, pp.4379--4382, 2019. 

\bibitem{17}L. Zou, X. Ding, Y. Zou, et al., "High power all-solid-state quasi-continuous-wave tunable Ti:sapphire laser system", \textit{Chin. Opt. Lett.}, Vol. 3, pp. 208-209, 2005.

\bibitem{18}Y. Li, D. Huang, H. Chen, et al., "Phase Noise of Fourier Domain Mode Locked Laser Based Coherent Detection Systems", \textit{J. Lightwave Technol.}, vol. 40, no. 3, pp. 615--623, 2022.

\bibitem{19}M. Kim, S. Ahn, J. Kim, et al., "Ultra-wideband wavelength-swept laser with a 440nm scanning range using four SOAs", \textit{29th International Conference on Optical Fiber Sensors}, vol. 13639, no. 136395S, 2025.

\bibitem{20}Y. Shi; D. Huang; Y. Li, et al., "High-Speed Multi-Gas Sensing With a Broadband Fourier Domain Mode-Locked Laser", \textit{J. Lightwave Technol.}, vol. 43, no. 6, pp. 2936--2942, 2025.	

\bibitem{21}C. Franken, R. Cheng, K. Powell, et al., "High-power and narrow-linewidth laser on thin-film lithium niobate enabled by photonic wire bonding", \textit{APL Photonics}, vol. 10, no. 026107, 2025.

\bibitem{22}C. Yang, Y. Li, Q. Cui, et al., "Narrow-linewidth, high-power, widely-tunable III-V/Si3N4 hybrid integrated external cavity laser", \textit{Opt. Laser Technol.}, vol. 186, no. 112627, 2025.

\bibitem{23}J. Fuchsberger, T. Letsou, D. Kazakov, et al., "Continuously and widely tunable semiconductor ring lasers", \textit{Optica}, vol. 12, pp. 985--990, 2025.

\bibitem{24}Liu Y., Chen Y., Bogaert L., Soltanian E., et al., "Widely tunable narrow-linewidth lasers with booster amplification on silicon photonics", \textit{Opt. Express}, vol. 33, no. 22078--22086, 2025.

\bibitem{25}Wang, Y.; Song, Y. et al., "Recent Advances in Tunable External Cavity Diode Lasers.", \textit{Appl. Sci.}, vol. 15, no. 206, 2025.

\bibitem{26}Zhao P., Shekhawat V., Girardi M. et al., "Ultra-broadband optical amplification using nonlinear integrated waveguides", \textit{Nature}, vol. 640, pp. 918--923, 2025.

\bibitem{27}M. Zou, K. Shen, Q. Song, et al., "Sub-kHz-linewidth laser generation by self-injection locked distributed feedback fiber laser", \textit{Opt. Laser Technol.}, vol. 169, no. 110022, 2024. 

\bibitem{28}X. Li, Y. Guo, E. Siyu, et al., "Ultra-narrow-linewidth high-power hybrid integrated self-injection locking laser on the silicon platform", \textit{2024 Conference on Lasers and Electro-Optics Pacific Rim (CLEO-PR)}, PDP\_8, 2024.

\bibitem{29}W. Liang, V. S. Ilchenko, A. A. Savchenkov, et al., "Whispering-gallery-mode-resonator-based ultranarrow linewidth external-cavity semiconductor laser", \textit{Opt. Lett.}, vol. 35, pp. 2822--2824, 2010.

\bibitem{30}K. Liu, N. Chauhan, J. Wang, et al., "36 Hz integral linewidth laser based on a photonic integrated 4.0 m coil resonator", \textit{Opitca}, vol. 9, pp. 770--775, 2022.

\bibitem{31}Idjadi M. H., Aflatouni F., "Mode-hop-free wavelength-tunable semiconductor lasers covering the S+C+L band via external-cavity optical feedback", \textit{Nat. Commun}, vol. 8, no. 1209, 2017.

\bibitem{32}I. Kabakova, R. Pant, D. Choi, et al., "Narrow linewidth Brillouin laser based on chalcogenide photonic chip", \textit{Opt. Lett.}, vol. 38, pp. 3208--3211, 2013.

\bibitem{33}M. Shtaif and G. Eisenstein, "Noise properties of nonlinear semiconductor optical amplifiers", \textit{Opt. Lett.}, vol. 21, pp. 1851--1853, 1996.

\bibitem{34}T. Zhu, X. Bao, L. Chen, et al., "Experimental study on stimulated Rayleigh scattering in optical fibers", \textit{Opt. Express}, vol. 18, pp. 22958--22963, 2010.

\bibitem{35}Y. Li, L. Huang, L. Gao, et al., "Optically controlled tunable ultra-narrow linewidth fiber laser with Rayleigh backscattering and saturable absorption ring", \textit{Opt. Express} vol. 26, pp. 26896--26906, 2018.

\bibitem{36}L. Dang, L. Huang, L.i Shi, et al., "Ultra-high spectral purity laser derived from weak external distributed perturbation", \textit{Opto-Electron. Adv.}, vol. 6 no. 210149, 2023.

\bibitem{37}D. Wei; L. Shi; J. Li, "Narrow Linewidth 1064 nm Laser Diode With External Distributed Feedback", \textit{J. Lightwave Technol.}, vol. 42, no. 24, pp. 8787-8792, 2024.

\bibitem{38}J. Li, D. Wei, L. Shi, et al., "Tunable narrow linewidth DFB laser diode with artificially enhanced Rayleigh scattering-based external distributed feedback", \textit{Opt. Express}, vol. 32, pp. 43771-43777, 2024.

\bibitem{39} K. Petermann, "Laser Diode Modulation and Noise", Kluwer Academic, Dordrecht, 1988.

\bibitem{40}C. Zhang, T. Guan, L. Huang, et al., "Adaptive distributed-feedback semiconductor laser", \textit{Phys. Rev. A}, vol. 111, no. 043511 , 2025.

\bibitem{41}C. Henry, "Theory of spontaneous emission noise in open resonators and its application to lasers and optical amplifiers", \textit{J. Lightwave Technol.}, vol. 4, no. 3. pp. 288--297, 1986.

\bibitem{42}N. Schunk and K. Petermann, "Noise analysis of injection-locked semiconductor injection lasers", \textit{ IEEE J. Quantum Electron.}, vol. 22, no. 5, pp. 642--650, 1986.

\bibitem{43}Y. Li, L. Huang, L. Gao, et al., "Optically controlled tunable ultra-narrow linewidth fiber laser with Rayleigh backscattering and saturable absorption ring", \textit{Opt. Express}, vol. 26, pp. 26896--26906, 2018.

\bibitem{44}L. Dang, L. Huang, Y. Cao, et al., "Side mode suppression of SOA fiber hybrid laser based on distributed self-injection feedback", \textit{Opt. Laser Technol.}, vol. 147, no. 107619, 2022.

\bibitem{45}R. Kumar, G. Su, D. Huang, et al., "Fully Integrated Tunable III-V/Si Laser With On-Chip SOA", \textit{J. Lightwave Technol.}, vol. 42, no. 9, pp. 3314--3319, 2024.

\bibitem{46}A. Malik, J. Guo, M. A. Tran, et al., "Widely tunable, heterogeneously integrated quantum-dot O-band lasers on silicon", \textit{Photonics Res.}, vol. 8, no. 1551, 2020.

\bibitem{47}N. Kobayashi, K. Sato, M. Namiwaka, et al., "Silicon Photonic Hybrid Ring-Filter External Cavity Wavelength Tunable Lasers", \textit{J. Lightwave Technol.}, vol. 33, no. 6, pp. 1241--1246, 2015.

\bibitem{48}H. Al-Taiy, N. Wenzel, S. Preußler, et al., "Ultra-narrow linewidth, stable and tunable laser source for optical communication systems and spectroscopy", \textit{Opt. Lett.}, vol. 39, pp. 5826--5829, 2014.

\bibitem{49}R. Zhou, S. Latkowski, J. O’Carroll, et al., "40nm wavelength tunable gain-switched optical comb source", \textit{Opt. Express}, vol. 19, pp. B415--B420, 2011.

\bibitem{50}Y. Gao, J. Lo, S. Lee, et al., "High-Power, Narrow-Linewidth, Miniaturized Silicon Photonic Tunable Laser With Accurate Frequency Control", \textit{J. Lightwave Technol.}, vol. 38, no. 2, pp. 265--271, 2020.

\bibitem{51}S. Bennetts, G. D. McDonald, K. S. Hardman, et al., "External cavity diode lasers with 5kHz linewidth and 200 nm tuning range at 1.55 $\mu$m and methods for linewidth measurement", \textit{Opt. Express}, vol. 22, pp. 10642-10654, 2014.

\bibitem{52}A. Becker, V. Sichkovskyi, M. Bjelica, et al., "Widely tunable narrow-linewidth 1.5 $\mu$m light source based on a monolithically integrated quantum dot laser array", \textit{Appl. Phys. Lett.}, vol. 110, no. 181103, 2017.

\bibitem{53}F. Aflatouni and H. Hashemi, "Wideband tunable laser phase noise reduction using single sideband modulation in an electro-optical feed-forward scheme", \textit{Opt. Lett.}, vol. 37, pp. 196--198, 2012.

\end{thebibliography}
\end{document}